\documentclass[pra,aps,onecolumn,superscriptaddress,showpacs,nofootinbib]{revtex4-2}

\usepackage[dvips]{graphicx}
\usepackage{amsmath,amssymb,amsthm,mathrsfs,amsfonts,dsfont}
\usepackage{subfigure, epsfig}
\usepackage{physics}
\usepackage{bm}
\usepackage{enumerate}
\usepackage{color}
\usepackage{qcircuit}
\usepackage[normalem]{ulem} % some line form: delete line
\usepackage{epstopdf}
\usepackage{comment}
\usepackage{diagbox}
\usepackage{multirow}    % multiculumn multirow
\usepackage{array}     % column centering
\usepackage{enumitem} % for item without label
\usepackage{booktabs,colortbl} % for long table
\usepackage{rotating}
\usepackage{ulem}
\usepackage{longtable}
\usepackage{setspace}
\usepackage[ruled,vlined]{algorithm2e}
\usepackage[colorlinks = true]{hyperref}
\SetKwInput{KwInput}{Input}
\SetKwInput{KwOutput}{Output}

\graphicspath{{./Figure/}}
\begin{document}
	\preprint{APS/123-QED}

\title{Field test of mode-pairing quantum key distribution}

\author{Hao-Tao Zhu}
\affiliation{Hefei National Research Center for Physical Sciences at the Microscale and School of Physical Sciences, University of Science and Technology of China, Hefei 230026, China}
\affiliation{CAS Center for Excellence in Quantum Information and Quantum Physics, University of Science and Technology of China, Hefei, Anhui 230026, China}
\affiliation{Hefei National Laboratory, University of Science and Technology of China, Hefei, Anhui 230088,China}

\author{Yizhi Huang}
\affiliation{Center for Quantum Information, Institute for Interdisciplinary Information Sciences, Tsinghua University, Beijing 100084, China}

\author{Wen-Xin Pan}
\author{Chao-Wu Zhou}
\affiliation{Hefei National Research Center for Physical Sciences at the Microscale and School of Physical Sciences, University of Science and Technology of China, Hefei 230026, China}
\affiliation{CAS Center for Excellence in Quantum Information and Quantum Physics, University of Science and Technology of China, Hefei, Anhui 230026, China}
\affiliation{Hefei National Laboratory, University of Science and Technology of China, Hefei, Anhui 230088,China}

\author{Jianjun Tang}
\affiliation{China Telecom Research Institute, Beijing 102209, China}
\author{Hong He}
\affiliation{China Telecom Co., LTD. Zhejiang branch, Hangzhou 310000, China}
\author{Ming Cheng}
\affiliation{China Telecom Research Institute, Shanghai 200122, China}
\author{Xiandu Jin}
\affiliation{China Telecom Co., LTD. Zhejiang branch, Hangzhou 310000, China}

\author{Mi Zou}
\affiliation{Hefei National Research Center for Physical Sciences at the Microscale and School of Physical Sciences, University of Science and Technology of China, Hefei 230026, China}
\affiliation{CAS Center for Excellence in Quantum Information and Quantum Physics, University of Science and Technology of China, Hefei, Anhui 230026, China}
\affiliation{Hefei National Laboratory, University of Science and Technology of China, Hefei, Anhui 230088,China}

\author{Shibiao Tang}
\affiliation{QuantumCTek Corporation Limited, Hefei, Anhui 230088, China}

\author{Xiongfeng Ma}
\email{xma@tsinghua.edu.cn}
\affiliation{Center for Quantum Information, Institute for Interdisciplinary Information Sciences, Tsinghua University, Beijing 100084, China}

\author{Teng-Yun Chen}
\email{tychen@ustc.edu.cn}
\author{Jian-Wei Pan}
\email{pan@ustc.edu.cn}
\affiliation{Hefei National Research Center for Physical Sciences at the Microscale and School of Physical Sciences, University of Science and Technology of China, Hefei 230026, China}
\affiliation{CAS Center for Excellence in Quantum Information and Quantum Physics, University of Science and Technology of China, Hefei, Anhui 230026, China}
\affiliation{Hefei National Laboratory, University of Science and Technology of China, Hefei, Anhui 230088,China}

\begin{abstract} 
Quantum key distribution is a cornerstone of quantum technology, offering information-theoretical secure keys for remote parties. With many quantum communication networks established globally, the mode-pairing protocol stands out for its efficacy over inter-city distances using simple setups, emerging as a promising solution. 
In this study, we employ the mode-pairing scheme into existing inter-city fiber links, conducting field tests across distances ranging from tens to about a hundred kilometers. Our system achieves a key rate of $1.217$ kbit/s in a $195.85$ km symmetric link and $3.089$ kbit/s in a $127.92$ km asymmetric link without global phase locking. The results demonstrate that the mode-pairing protocol can achieve key rates comparable to those of a single quantum link between two trusted nodes on the Beijing-Shanghai backbone line, effectively reducing the need for half of the trusted nodes. These field tests confirm the mode-pairing scheme's adaptability, efficiency, and practicality, positioning it as a highly suitable protocol for quantum networks.	
\end{abstract}
\maketitle
%%%%%%%%%%%%%%%%%%%%%%%%%%  body  %%%%%%%%%%%%%%%%%%%%%%%%%%
\section{Introduction}
Quantum key distribution (QKD), standing at the forefront of quantum technology, has captured significant attention for its unique capability to securely distribute cryptographic keys, impervious to even the most powerful eavesdroppers with unlimited computational powers \cite{bennett1984quantum,ekert1991Quantum}. Beginning with the first proof-of-concept demonstrations~\cite{Bennett1992experimental}, QKD has rapidly evolved into sophisticated and practical implementations. A landmark achievement was the deployment of satellite-based QKD, enabling intercontinental secure quantum communication~\cite{Liao2018Satellite}. This progress was furthered by the construction of the Beijing-Shanghai backbone, a critical development towards a large-scale, ground-based quantum communication network, showcasing QKD's potential for national and global secure communication infrastructures~\cite{Chen2021integrated}. These developments have not only demonstrated the practicality of QKD but also set the stage for integrating quantum communication into mainstream networks.

Following initial achievements in QKD, the development of measurement-device-independent (MDI) QKD marked a significant stride forward, addressing critical vulnerabilities associated with detector side-channel attacks. Introduced by Lo, Curty, and Qi in 2012~\cite{lo2012Measurement}, MDI QKD effectively decouples the security of quantum communication from the trustworthiness of the detection apparatus. The primary strength of MDI QKD lies in its inherent immunity to all forms of detector-side attacks. Subsequent implementations and experiments have demonstrated MDI QKD's practicality and scalability, paving the way for more secure quantum networks~\cite{liu2013experimental,Tang2014Experimental}.

The implementation of MDI QKD in quantum networks introduces a significant advantage by allowing the inclusion of many untrusted nodes. This reduces maintenance costs for the network and offers greater flexibility in node placement. However, the performance of MDI QKD is limited by the linear relationship between key rate and channel transmittance \cite{takeoka2014fundamental,pirandola2017fundamental}, which hampers long-distance communication or achieving high key rates in intercity quantum networks. Consequently, this poses a practical barrier to extending the reach and efficiency of quantum networks over large distances.

The evolution of MDI QKD protocols further advanced with the twin-field scheme \cite{lucamarini2018overcoming}, enabling higher key rates over extended distances and effectively overcoming the rate-distance limit of conventional QKD~\cite{Ma2018phase,lin2018simple}. Subsequently, the phase-matching scheme emerged as a variant of the twin-field scheme, offering rigorous security proof and further improving the performance in real-world fiber networks~\cite{Ma2018phase,Zeng2019Symmetryprotected}. A key implementation challenge in the twin-field and phase-matching schemes is achieving global phase locking, a crucial requirement for maintaining the coherence between distant quantum states~\cite{minder2019experimental,fang2020implementation,chen2020sending,Mao2021recent,chen2021twin}. This requirement for precise phase stabilization adds complexity to its deployment, especially over long distances, where phase fluctuations are more pronounced, adds complexity to its deployment. The technical challenges in global phase locking and the additional consumption of optical fiber resources limit the practical application of this type of scheme in quantum networks.

Addressing the global phase-locking challenge, the mode-pairing (MP) scheme has emerged as an effective solution. Noted for its efficiency and simplified setup, the MP scheme bypasses the need for phase stabilization by leveraging the inherent stability in the pairing of modes~\cite{zeng2022mode}. See also Ref.~\cite{xie2022breaking} for a similar scheme without rigorous security proof. This approach maintains coherent relationships between quantum states without the stringent phase stabilization demands. Consequently, the MP protocol has shown remarkable performance over inter-city distances, making it a compelling and practical choice for quantum communication networks. The relative ease of its implementation, combined with its robustness, makes the MP scheme a promising approach for quantum communications.

The MP scheme has demonstrated its effectiveness in the laboratory, achieving impressive key generation rates~\cite{Zhu2023experimental,Zhou2023Experimental}. However, transitioning from controlled laboratory environments to real-world settings introduces new experimental challenges. Particularly, the performance of the MP scheme in field-deployed optical fiber networks and under asymmetric link conditions is yet to be verified. The complexities of field environments and the necessity of parameter optimization for multi-node scenarios to maximize key rates are critical considerations. These uncertainties underscore the need for rigorous field testing to validate the protocol's real-world applicability.

In this study, we conduct the MP scheme within a deployed inter-city optical fiber network. Our experiments are designed for symmetric and asymmetric link scenarios, aiming to assess the protocol's performance under real-world conditions comprehensively. The field-test experiment is performed in a typical scale of existing inter-city quantum networks, such as the Beijing-Shanghai backbone link \cite{Chen2021integrated}. We employed a linear programming approach for parameter estimation to improve the key rate, especially in the asymmetric scenario. By utilizing commercial fibers with ultra-physical contact (UPC) connectors and integrating fiber circulators, our field test effectively mitigates the impact of light reflecting from detectors and the UPC fiber end face. This approach significantly reduces noise, achieving an exceptionally low bit error rate in the order of parts per ten thousand. Notably, our demonstration avoids using global phase-locking techniques, simplifying the system configuration and conserving additional fiber resources. The elimination of global phase-locking contributes to reduced background noise and more stable error rates, enabling the implementation of only two dense wavelength division multiplexing (DWDM) filters for noise reduction. This further streamlines the system and enhances its practicality.

Our results highlight the performance of our system, achieving a key rate of several thousand bits per second within a communication range of 100 to 200 kilometers in both symmetric and asymmetric scenarios. This key rate proves comparable to that of a single QKD system, even over distances surpassing the average inter-node spans in the Beijing-Shanghai backbone network. Underscoring the practical viability of the MP scheme, with its straightforward design akin to operational networks, these findings show its flexibility and superior key generation rates compared to existing QKD schemes, all while preserving MDI characteristics. The MP scheme's adaptability to urban quantum communication networks enhances node placement flexibility and routing capabilities, offering a robust solution for diverse network scenarios. Demonstrating its suitability for real-world inter-city quantum networks, the MP scheme significantly improves secure communication by offering enhanced key generation rates without the need for trusted node deployments. 

This paper is organized as follows. In Section \ref{sc:SchemeDescrip}, we briefly introduce the MP scheme. In Section \ref{sc:setup}, we detail the experimental setup. In Section \ref{sc:results}, we present the results and discussions. In appendixes, we provide more details of the scheme and setting, as well as test results of the system and experimental raw data.

\section{Mode-Pairing Scheme} \label{sc:SchemeDescrip}
In this section, we introduce the MP scheme utilized in our field test, primarily based on the theoretical framework outlined in \cite{zeng2022mode}. However, there are several critical modifications tailored for real-world field testing. Firstly, due to the often asymmetric nature of Alice's and Bob's channels in field environments, they opt for different parameters for their encoding strategies. Secondly, due to the asymmetry in the communication channel leading to different intensities sent by Alice and Bob, the analytical formulas used in the original MP scheme for estimating the single-photon component and phase error rate are no longer applicable. In practical experiments, we employed a linear programming approach for parameter estimation. Moreover, to effectively estimate phase fluctuations in the channel, we incorporate the use of strong light pulses as reference in the protocol and use the maximum-likelihood estimation method to estimate the phase differences with the detection results of the strong light pulses. 
%[xma: others?] 
The protocol encompasses the following steps, with details in Appendix \ref{app:schemedetail}.

%periods duration; MLE; The strategies for pairing and basis assignment used in our experiments are consistent with those detailed in \cite{Zhu2023experimental}.

\emph{State Preparation:} Alice and Bob divide a session into three periods: strong light, recovery, and signal light. In the strong light period, they send strong light pulses without phase encoding, with intensities set to ensure sufficient clicks at the detection site. The recovery period involves no pulse transmission, aimed at eliminating the effects of the strong light pulses. In the signal light period, Alice generates a coherent state $\ket{\alpha}$, where $\alpha=\abs{\alpha}e^{\mathrm{i} \phi^a}$. She chooses its intensity $\abs{\alpha}^2$ randomly from the set $\{0,\nu_a,\mu_a\}$ and its phase $\phi^a$ from $\{0, \frac{2\pi}{D}, \ldots, \frac{2\pi(D-1)}{D}\}$. Similarly, Bob generates a coherent state $\ket{\beta}$, where $\beta=\abs{\beta}e^{\mathrm{i} \phi^b}$, with intensity from $\{0,\nu_b,\mu_b\}$ and phase from the same set as Alice's. Note that to accommodate potential asymmetries in the communication channels, Alice and Bob may opt for different intensity sets and selection probabilities.

\emph{State Transmission and Measurement:} Alice and Bob send their prepared pulses to the measurement site managed by Charlie, who then performs single-photon interference measurements and announces detector clicks: $L$ or $R$.

\emph{Frequency Estimation:} 	Alice and Bob repeat the state preparation and transmission steps until each has sent a total of $N$ pulses during the signal light periods. Subsequently, they engage in post-processing of the data. From the data generated in the strong pulse periods, they estimate the underlying frequency differences arising from the lasers and optical fibers, employing either the maximum likelihood estimation method \cite{Zhu2023experimental} or the Fast Fourier Transform Algorithm \cite{li2023twin}.

\emph{Pairing and Basis Assignment:} For data generated in the signal pulse periods, Alice and Bob employ a predetermined strategy for pairing rounds with successful detection outcomes according to the maximum pairing length $L_{max}$. They announce the sum of intensities and the assigned basis for each pair. This process then enables them to sift the $X$-pairs and $Z$-pairs. Considering the potential asymmetry in the intensities of Alice's and Bob's pulses, the total intensities $\vec{\mu}=(\mu_a^i+\mu_a^j,\mu_b^i+\mu_b^j)$ for paired rounds $i$ and $j$ are selected to fulfill specific criteria. For the $Z$-pairs, $\vec{\mu}$ falls within the set $\{\vec{\mu}_z\} \equiv \{(\mu_a,\mu_b),(\mu_a,\nu_b),(\nu_a,\mu_b),(\nu_a,\nu_b)\}$. For the $X$-pairs, $\vec{\mu}$ adheres to the set $\{\vec{\mu}_x\} \equiv \{(2\mu_a,2\mu_b),(2\mu_a,2\nu_b),(2\nu_a,2\mu_b),(2\nu_a,2\nu_b)\}$.

\emph{Key Mapping:} Alice and Bob proceed to map raw key bits from the intensity and phase information of the pairs. For each $Z$-pair, Alice sets her key bit $\chi_a=0$ if the intensity of the $i$-th pulse is $\mu_i^a=0$ and $\chi_a=1$ if $\mu_j^a=0$. Bob assigns $\chi_b=1$ if $\mu_i^b=0$ and $\chi_b=0$ if $\mu_j^b=0$. For each $X$-pair, Alice extracts her key bit $\chi_a$ based on the relative phase, $\chi_a=\lfloor(\phi^a_j-\phi^a_i)/\pi \mod 2\rfloor$, and then announces $\theta_a=(\phi^a_j-\phi^a_i)\mod \pi$. Bob follows a similar procedure for assigning his raw key bits $\chi_b$ and announces $\theta_b$. Alice and Bob retain the key only if 
\begin{equation}
	|\theta_b-\theta_a| \in \left[ \Delta\theta_{i,j}+2k\pi-\frac{\pi}{D},\Delta\theta_{i,j}+2k\pi+\frac{\pi}{D}\right],
\end{equation}
where $k$ is an integer, $\Delta\theta_{i,j} = \int_{t_i}^{t_j}\Delta\omega(t) dt$, $\Delta\omega(t)$ represents the estimated frequency difference, and ${t_i},{t_j}$ are the corresponding times of the $i$-th and $j$-th rounds. Keys are discarded if this condition is not met.

\emph{Parameter Estimation:} Alice and Bob use $Z$-pairs with different intensity settings to estimate the count of clicked single-photon pairs $M^Z_{11}$, applying the decoy-state method \cite{hwang2003decoy,Lo2005Decoy,wang2005decoy}. For $X$-pairs, they estimate the single-photon phase-error rate $e^{Z,ph}_{11}$ \cite{zeng2022mode}. They also employ the Chernoff-Hoeffding method to account for statistical fluctuations \cite{Zhang2017improved}. 

\emph{Error Correction and Privacy Amplification:} Alice and Bob generate raw key bits using $Z$-pairs, resulting in $M_{\vec{\mu}}$ bits. They perform error correction based on the quantum bit error rate $E_{\vec{\mu}}$ and then execute privacy amplification based on the estimated parameters $M^Z_{11}$ and $e^{Z,ph}_{11}$.

Finally, Alice and Bob can extract a secure key of length
\begin{equation}\label{eq:keyR}
	K =M^{Z}_{11} \left[1 - h\left(e^{Z,ph}_{11}\right)\right] -\sum_{\vec{\mu}\in\{\vec{\mu}_z\}} f M_{\vec{\mu}} h\left(E_{\vec{\mu}} \right),
\end{equation}
where $h(x) = -x \log_2 x - (1-x)\log_2(1-x)$ represents the binary entropy function, and $f$ denotes the error correction efficiency. 
%The key rate $R$ is defined as the key length $K$ divided by half the total number of rounds $N/2$.

\section{Experimental Description}\label{sc:setup}
\begin{figure}[h]
	\centering
	\includegraphics[width=0.8\textwidth]{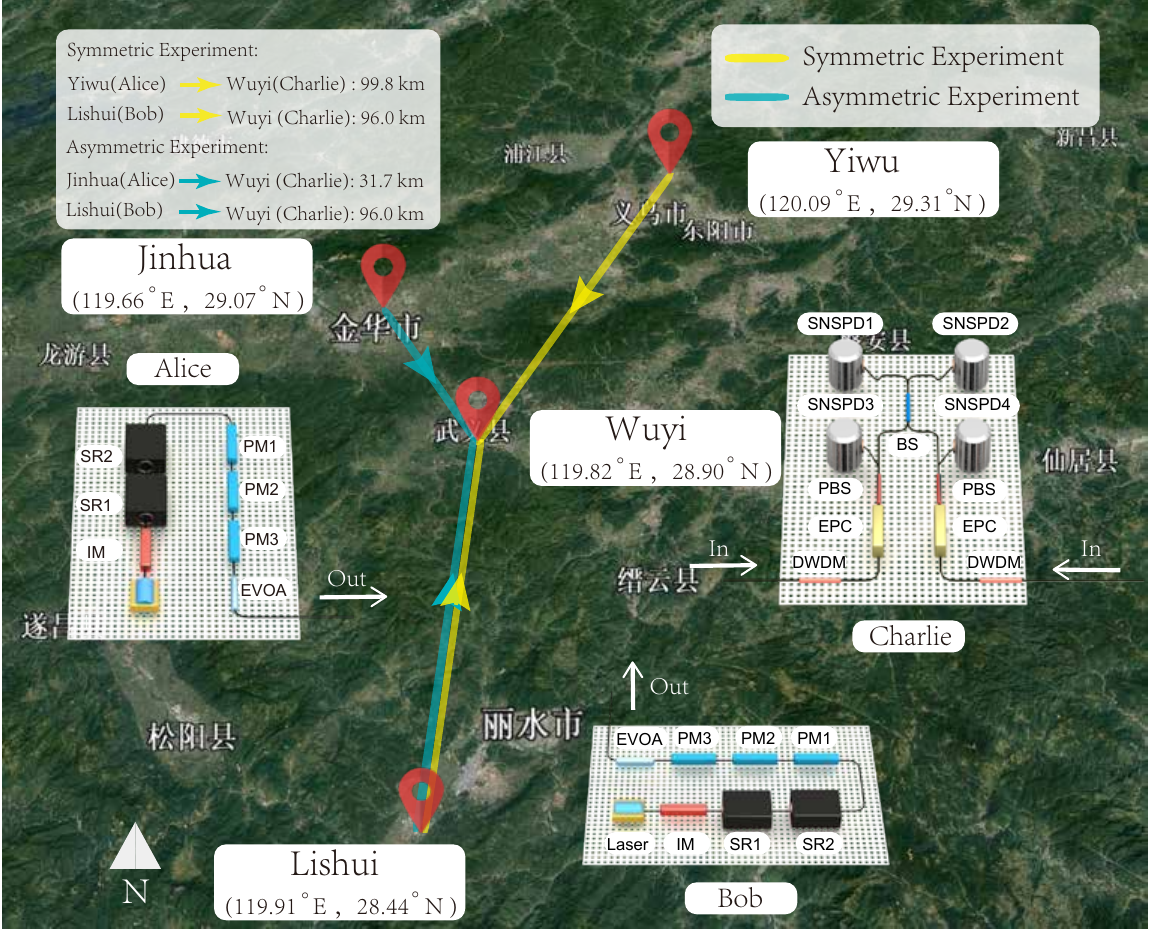}
	\caption{Experimental setup. The measurement site is situated in a telecommunication room in Wuyi (119°49'47" E, 28°54'26" N). The senders operate from telecommunication rooms in Yiwu (120°5'40" E, 29°18'52" N), Lishui (119°54'58" E, 28°26'55" N), and Jinhua (119°39'39''E, 29°4'15''N). Map data sourced from Google, Landsat/Copernicus. The configuration is tailored for both symmetric and asymmetric experiments. Two independent lasers, housed in separate transmitter rooms, emit continuous light, subsequently chopped into pulses at a frequency of 625 MHz by an intensity modulator. These pulses undergo modulation by two Sagnac rings, followed by an additional modulation with 16 distinct random phases using three phase modulators. The pulse intensity is then attenuated to the single-photon level through an EVOA before transmission to Charlie via buried optical fiber. At Charlie, DWDM is utilized to filter out noise beyond the quantum optical wavelength. Polarization feedback for Alice and Bob is achieved through EPC, PBS, SNSPD3, and SNSPD4, while interference measurements involve BS, SNSPD1, and SNSPD2. Abbreviations:	
		IM: intensity modulator, SR: Sagnac ring, PM: phase modulator, EVOA: electric variable optical attenuator, DWDM: dense wavelength division multiplexer, EPC: electric polarization controller, PBS: polarization beam splitter, SNSPD: superconducting nanowire single photon detector, BS: 50:50 beam splitter.% FIBRE: commercial fiber,
	}
	\label{fig:ExpSetup}
\end{figure}

In this section, we present a comprehensive overview of the experimental setup employed in our field test for the MP QKD experiment. Figure \ref{fig:ExpSetup} offers an aerial perspective of our field test setup, showcasing the configuration at each site. Our experiment is conducted using four data center nodes strategically positioned within a commercial optical fiber communication network. These nodes, located in the cities of Wuyi, Lishui, Jinhua, and Yiwu, are chosen for our communication and measurement needs. The central node in Wuyi, noted as Charlie, functions as the measurement terminal, while the remaining nodes serve as senders, labeled as Alice and Bob.

The communication terminals establish connections with the measurement terminal through pre-existing commercial optical fiber cables, creating three communication links with varying lengths, as illustrated in Figure \ref{fig:ExpSetup}. To assess symmetric channel performance, we select the Lishui-Wuyi-Yiwu route, where the distances are roughly equal.  For the asymmetric channel test, we selected the Lishui-Wuyi-Jinhua route, introducing variations in communication distances. The specific parameters for these routes are consolidated in Table \ref{table:Fieldparameters}. Here, the detection efficiency is denoted as $\eta_{det}$, and $\epsilon$ represents the security parameter.

The choice of these routes is deliberate and considers various factors. Firstly, in terms of distances, they mirror typical separations between cities, providing a basis for evaluating the effectiveness of the MP scheme in intercity quantum communication scenarios. Additionally, the considerable difference in distances between the communication parties and the measurement terminal in the asymmetric route demonstrate the notable flexibility and scalability of the MP scheme when applied in quantum network scenarios.

\begin{table}[h]
	\caption{Experimental parameters.}
	\centering
	\begin{tabular}{l|l|l|l|l|l|l|l|l	}%{p{2cm}|p{2cm}|p{2cm}|p{2cm}|p{2cm}|p{2cm}|p{2cm}|p{2cm}}
		\hline
		\hline
		& $ L_{A-C} $ & $ L_{B-C} $ & $ \eta_{A-C} $ & $ \eta_{B-C} $ & $ \eta_{det} $ &$L_{max}$& $f$ & $\epsilon$ \\
		\hline
		Symmetry & 99.84\;km & 96.01\;km & $ 5.767\times10^{-3} $ & $ 6.004\times10^{-3} $ & 72 $\%$ &1000& 1.06 & $ 10^{-10} $ \\		
		\hline
		Asymmetry & 31.91\;km & 96.01\;km & $ 7.523\times10^{-2} $ & $ 5.723\times10^{-3} $ & 72 $\%$ & 1000&1.06 & $ 10^{-10} $ \\		
		\hline
		\hline
	\end{tabular}	
	\label{table:Fieldparameters}
\end{table}

In Figure \ref{fig:ExpSetup}, we present a detailed overview of our experimental setup across various sites. Alice and Bob emit continuous light from a narrow-linewidth laser with a central wavelength of 1550.12 nm and a 2 kHz linewidth. This light undergoes chopping into pulses, with its intensity and phase modulated using an electrical signal at 625 MHz generated by the Field-Programmable Gate Array (FPGA). Further, the continuous light is chopped into pulses at 625 MHz using the intensity modulator (IM), with the QKD working frequency set at 444.83 MHz. Reference pulses come into play for estimating the frequency difference between the two lasers, as elaborated in Appendix \ref{app:schemedetail}. Subsequently, having traversed two Sagnac rings (SRs), four types of pulse light are modulated: reference, signal, decoy, and vacuum pulses. The introduction of three additional phase modulators (PMs) serves to randomize the phase into 16 separated phases, ranging from ${0, (2\pi)/16, (4\pi)/16, \cdots, 30\pi/16}$. Following this phase modulation, pulses are attenuated to the single-photon level and transmitted through optical fiber links to Charlie for interference measurement.

At the measurement site, Charlie employs DWDMs with a bandwidth of 100 G to effectively filter out light noise from other wavelengths. Additionally, Charlie utilizes an electric polarization controller, a polarization beam splitter, and superconducting nanowire single photon detectors (SNSPDs), specifically SNSPD3 and SNSPD4, for real-time polarization feedback. For interference detection, a beamsplitter and two additional SNSPDs, SNSPD1 and SNSPD2, are employed. The detector efficiency is measured at 72\%, with a dark count rate of 36.5 Hz. Clock synchronization for the three separated nodes is accomplished by transmitting synchronization light through another optical fiber. In the synchronization process, Charlie utilizes 1570 nm light pulses at 100 kHz during both symmetric and asymmetric field tests. These pulses are divided by a beamsplitter and transmitted to Alice and Bob. In their respective locations, Alice and Bob use a high-bandwidth positive-intrinsic-negative diode to convert the light signal into an electrical signal. The 100 kHz electrical signal triggers the FPGA for generating a 625 MHz synchronized electrical signal, which subsequently drives the intensity and phase modulators of pulse modulation.

In our field tests, we employ commercial fiber with UPC connectors, requiring the inclusion of two fiber circulators in Charlie's setup. This configuration is designed to prevent light pulses from reflecting back into the detector after being reflected by the UPC fiber end face, thereby avoiding the introduction of noise into our systems. The loss introduced by the two fiber circulators is taken into account in Charlie's optical setup, contributing to low noise levels and ultimately yielding a $Z$-basis error rate as minimal as a few parts per ten thousand.

\begin{figure}[h]
	\centering
	\includegraphics[width=0.7\textwidth]{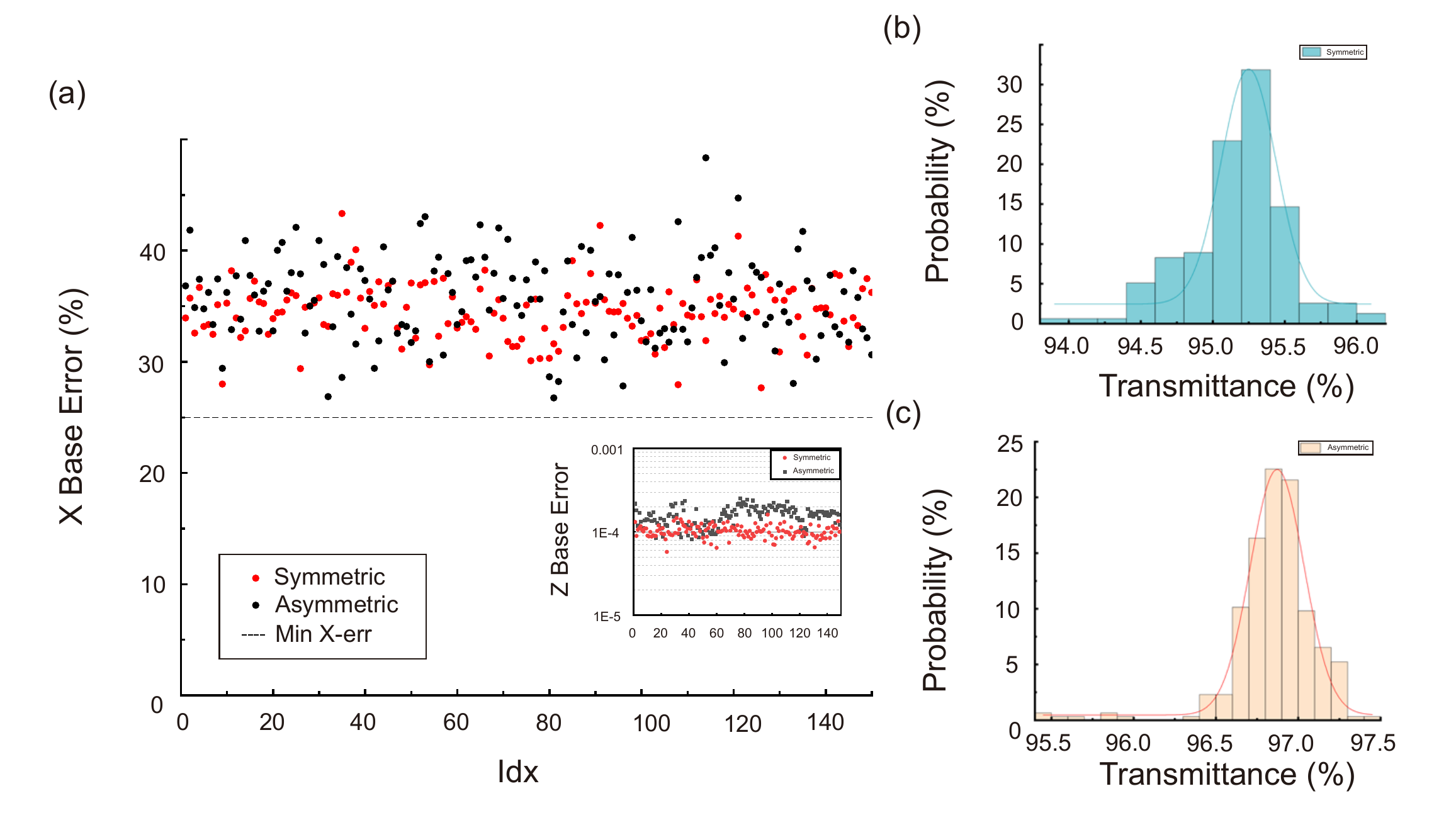}
	\caption{
		Error rate drift and polarization feedback efficiency. (a) Error rates in the $X$-basis and $Z$-basis. Each data point represents the effective clicks collected over 30 seconds for the symmetric scenario and 15 seconds for the asymmetric scenario. (b) Probability distribution of transmittance for the PBSs in Charlie during the symmetric test, with a total efficiency of 95.19\%. (c) Probability distribution of transmittance for the PBSs in Charlie during the asymmetric experiment, with a total efficiency of 96.87\%.	 }
	\label{fig:ErrorandPol}
\end{figure}

Unlike laboratory fiber spools, the field fiber environment is susceptible to increased interference, leading to unstable and elevated error rates. To evaluate this impact, we conducted tests on the disturbance of optical pulses' time patterns caused by the field fiber and observed variations in $Z$ and $X$ basis error rates over time, as depicted in Figure \ref{fig:ErrorandPol}. The test results emphasize that, due to the MP scheme's independence from phase-locking technology, the need for an additional laser and fiber resources for phase-locking is eliminated. This absence of an additional laser results in relatively low background noise and more stable error rates in both symmetric and asymmetric tests. As a result, only two DWDMs are required to effectively filter noise in the quantum channel, streamlining the entire system and enhancing its practicality. Additional results on channel stability tests are presented in Appendix \ref{app:stable}.

\section{Results and Conclusion}\label{sc:results}

Our main experimental results, summarized in Table~\ref{table:results}, reveal significant insights from the field tests under both symmetric and asymmetric scenarios. In each test, $2.00\times10^{12}$ pulses is transmitted by Alice (Bob). In the symmetric test, Alice and Bob obtain $88377993$ $(\mu_a,\mu_b)$-pairs for generating raw key bits, with an observed error rate of $0.01\%$. For the $X$ pairs, the error rate for the $(2\mu_a,2\mu_b)$-pair is recorded at $33.92\%$. Considering the finite data size effect, the single-photon component of the $Z$-basis pairs, denoted as $M_{11}^Z$, is lower-bounded to be $43004502$, with a corresponding phase error rate $e^{Z,ph}_{11}$ upper-bounded by $31.31\%$. This results in a final key rate per pair in the symmetric case of $5.47\times10^{-6}$, corresponding to an effective key rate of $1.217$ kbit/s.

\begin{table}[h]
	\caption{Experimental results. Symmetric route: Lishui-Wuyi-Yiwu and asymmetric route: Lishui-Wuyi-Jinhua. }
	\centering
	\begin{tabular}{l|l|l|l|l|l|l}
		\hline
		\hline
		Route	& $ N $ & $M_{(\mu_a,\mu_b)} $ & $E_{(\mu_a,\mu_b)}$& $e^{Z,ph}_{11}$ & $R$ (bit/pair) & $R$ (bit/s) \\
		\hline
		Symmetric &$2.00\times10^{12}$ & 88377993 & $ 0.01\%$ & $31.31\%$ & $5.47\times10^{-6}$ & $1217.17$ \\		
		\hline
		Asymmetric & $2.00\times10^{12}$ & 294929141& $ 0.02\%$ & $31.52\%$ & $1.38\times10^{-5}$ & $3088.70$\\		
		\hline
		\hline
	\end{tabular}	
	\label{table:results}
\end{table}

In the asymmetric test among Lishui, Wuyi, and Jinhua, Alice and Bob successfully collect $294929141$ $(\mu_a,\mu_b)$-pairs, with an error rate of $0.02\%$. The error rate for the $(2\mu_a,2\mu_b)$-pairs is $38.05\%$. With the decoy-state estimation, $M_{11}^Z$ is lower-bounded by $124813948$, with an associated $e^{Z,ph}_{11}$ upper-bounded by $31.52\%$. The final key rate per pair in this scenario is $1.38\times10^{-5}$, equating to an effective key rate of $3.089$ kbit/s. The presented data primarily focus on $(\mu_a,\mu_b)$-pair and $(2\mu_a,2\mu_b)$-pair, as they are the principal components of $Z$-pair and $X$-pair, respectively. For further details on experimental settings, results, and comparisons with simulations, readers are referred to the Appendix \ref{app:detaildata}.%Supplementary Material \cite{FieldTestMPsupp}.

Our experiments highlight the robust performance of the MP scheme in practical optical fiber environments. The system configuration in our study, akin to the one used in the Beijing-Shanghai backbone network \cite{Chen2021integrated}, features a comparably straightforward and implementable design. Importantly, our system achieves a key rate roughly equivalent to that of a single quantum key distribution system over distances exceeding the average inter-node distance in the Beijing-Shanghai backbone network. Notably, the implementation of wavelength-division multiplexing technology in the Beijing-Shanghai backbone network, which enables multiple systems to perform key generation simultaneously over a single optical fiber, could significantly enhance the key rates for the MP scheme by allowing pairing across different wavelength channels \cite{zeng2022mode}.

Moreover, compared to the current QKD schemes used in networks, the MP scheme maintains its MDI nature while offering a simple setup and high performance. In practical network applications, employing the MP scheme allows a certain amount of networks nodes to be untrusted, thus obviating the need for secure environments such as urban areas. This increases the potential distances between trusted nodes in a quantum network, substantially enhancing node placement flexibility. Additionally, the MP scheme is particularly well-suited for urban quantum communication networks, allowing nodes to function flexibly as either detection or source sites, which is advantageous for network routing and key management \cite{Zhou2022Network}.

Unlike the phase-matching scheme \cite{fang2020implementation}, the MP scheme eliminates the need for global phase locking. In field tests, it achieves key rates comparable to the twin-field protocol with global phase locking at distances around 200 km in field test \cite{liu2021field}. Our results also demonstrate that the key generation rate of the MP protocol substantially exceeds that of conventional measurement-device-independent protocols \cite{Tang2016MDInet} in real-world network settings. Moreover, our successful demonstration in asymmetric channels showcases the MP scheme's capability to maintain high key generation rates even under extreme asymmetric conditions. By optimizing appropriate parameters, we have demonstrated its adaptability to complex real-world scenarios.

In conclusion, the MP scheme presents an optimal solution that combines implementation flexibility, high performance, device simplicity, and operational ease. It simultaneously reduces the construction and maintenance costs of quantum networks, making it an excellent QKD protocol for quantum network applications.

%Backmatter sections should be listed in the order Funding/Acknowledgment/Disclosures/Data Availability Statement/Supplemental Document section. An example of backmatter with each of these sections included is shown below.

\acknowledgements
The authors acknowledge Lixing You for insightful discussions. This work has been supported by the National Natural Science Foundation of China Grant No.~12174216, the Innovation Program for Quantum Science and Technology Grant No.~2021ZD0300702 and No.~2021ZD0300804, and Anhui Initiative in Quantum Information Technologies.

\bibliographystyle{apsrev4-2}
\bibliography{bibMPExp.bib}

\appendix

\section{Protocol Details}\label{app:schemedetail}
Here, we provide a detailed account of several aspects of the MP protocol employed in our field test. This includes the sending period of different intensities in the state preparation step, the steps of the maximum likelihood estimation method, the procedures for pairing and basis assignment, finite-data-size analysis method, and finally how we utilize the decoy-state method to estimate the single-photon component $M_{11}^Z$ and its associated phase error rate $e_{11}^{Z,ph}$ in Eq.~\eqref{eq:keyR}.

\subsection{Sending period}
\begin{figure}[h]
	\centering
	\includegraphics[width=0.7\textwidth]{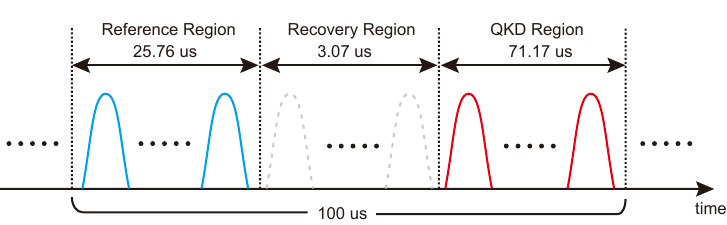}	
	\caption{Sending pulses sequence in one cycle. Alice and Bob periodically transmit pulses in cycles of 100 \textmu s. Within the reference region, no phase modulation is applied to the strong light pulses by Alice and Bob. Moving to the recovery region, pulses are modulated into vacuum states to mitigate the influence of strong light pulses. In the QKD region, Alice and Bob transmit pulses with varying intensities and phases. Subsequently, after interference, Charlie discloses the detected clicks from all detectors.} 	
	\label{fig:SendingPeriod}
\end{figure}

In Fig. \ref{fig:SendingPeriod}, we show the light pulses of Alice (Bob) sent in one cycle. The reference region is utilized for the two lasers' frequency difference estimation, which is introduced in Ref.~\cite{Zhu2023experimental}. The duration time of the reference region in one cycle is 25.76 \textmu s. The recovery region is designed to protect the pulses in the QKD region from the reference region's pulses. This region occupies 3.07 \textmu s. The pulses of the QKD region are used for secure key generation. The duty cycle of QKD is 71.17 \%.

\subsection{Maximum likelihood estimation method for phase estimation}
In the phase-estimation, the click results of strong light pulses are partitioned into distinct data groups, and the maximum likelihood estimation method is applied to each group. Practically, the time interval between the first and last clicks within the same data group is kept below 1 ms, allowing $\Delta\omega(t)$ to be considered a constant within a data group. To streamline the process, we omit the variable $t$ and consider $\Delta\omega$ as a constant within a single data group.

For the click data within a group, Alice and Bob pair arbitrary pairs of click events, such as the $i$th and $j$th clicks. Each click result, denoted as $D_i$ and $D_j$ for the $i$th and $j$th clicks, respectively, yields either an $L$-click (0) or an $R$-click (1). The probabilities of identical detector clicks, either $(0,0)$ or $(1,1)$ denoted as $P^0$, and clicks on each detector once, either $(0,1)$ or $(1,0)$ denoted as $P^1$, can be calculated as follows:
\begin{equation}
	\begin{aligned}
		P^0 &= \dfrac{1}{2}+\dfrac{\cos\Delta\theta}{4},\\
		P^1 &= \dfrac{1}{2}-\dfrac{\cos\Delta\theta}{4}.\\
	\end{aligned}
\end{equation}
Here, the phase difference $\Delta\theta$ between the two pulses is expressed as
\begin{equation}
	\Delta\theta = \Delta\omega\tau (j-i),
\end{equation}
where $\tau$ represents the time interval between two adjacent pulses.

Consequently, the likelihood function is defined as
\begin{equation}
	\begin{aligned}
		f(\Delta\omega) &=\ln \prod_{(i,j)} P^{|D_i-D_j|} \\
		&= \sum_{(i,j)} \ln\left\{\dfrac{1}{2}+(-1)^{D_i-D_j}\dfrac{\cos\left[\Delta\omega\tau (j-i)\right]}{4}\right\}. \\
	\end{aligned}
\end{equation}
The value of $\Delta\omega$ is then estimated by locating the maximum of the likelihood function within a predetermined range obtained from preliminary system tests, utilizing instruments like a frequency counter.

To obtain frequency differences for each data group, Alice and Bob iteratively follow the above steps. They fit the estimated results using the least square method to derive $\Delta \omega(t)$ for QKD pulses, employing a dataset consisting of 200 data groups. In practice, $\Delta \omega(t)$ exhibits gradual changes over time, as illustrated in Appendix \ref{app:stable}, prompting Alice and Bob to utilize the least square method for fitting.

\subsection{Pairing and basis assignment procedure}
In our field test, our pairing scheme follows the principle of pairing nearby pulses under the constraint of the maximum pairing length $L_{max}$. Specifically, Alice and Bob systematically scan all pulse response results in sequence, starting from the beginning and moving forward. Upon encountering the first pulse with successful detection, it is designated as the ``front pulse,'' and the scan continues. Upon finding the second pulse with successful detection, it is labeled as the ``rear pulse,'' and the distance between the front and rear pulses is calculated by subtracting their respective sequence numbers. If the distance is less than $L_{max}$, the two pulses are paired, and the scan progresses to the next front pulse. If the distance exceeds the maximum pairing length, the front pulse is discarded, and the current rear pulse becomes the new front pulse for the continued scan to identify subsequent rear pulses.

After the pairing process, Alice and Bob proceed to perform basis assignment for these pairs. Initially, each of them independently labels these pairs based on the intensities of the light pulses the two rounds. The labeling is carried out using Table~\ref{table:SingleSideBasis}, where, for paired $i$-th and $j$-th rounds, the corresponding intensities are denoted as $\mu^{a(b)}_i$ and $\mu^{a(b)}_j$, respectively.

\begin{table}[hbpt!] 
	\caption{Basis assignment and sifting depending on intensities.}\label{table:SingleSideBasis}
	\centering 
	\begin{tabular}{|c|ccc|}  
		\hline  
		\diagbox{$\mu^{a(b)}_i$}{$\mu^{a(b)}_j$} & 0 & $\nu$&$\mu$ \\   
		\hline   
		0		& `0' 	& $Z$	&$Z$ \\ 
		$\nu$ 	& $Z$ 	& $X$  	&`discard'\\  
		$\mu$	& $Z$ 	& `discard'& $X$  \\      
		\hline   
	\end{tabular} 
\end{table}

Subsequently, Alice and Bob publicly announce the labels of all pairs and the sum of the intensities $\vec{\mu}\equiv(\mu_i^a+\mu_j^a,\mu_i^b+\mu_j^b)$ for each pair. They further refine the basis of these pairs according to Table~\ref{table:TwoPartyBasis}. Note that the ``discard'' entries in Table~\ref{table:SingleSideBasis} have no impact on subsequent data processing and can be directly discarded. Therefore, cases where one side is labeled as "discard" are not explicitly listed here. 

\begin{table}[hbpt!]
	\caption{Basis sifting according to intensity information from both sides.} \label{table:TwoPartyBasis}
	\centering 
	\begin{tabular}{|c|ccc|}   
		\hline   
		\diagbox{Alice}{Bob} & `0' & $Z$&$X$ \\   
		\hline   
		0		& `0' 	& $Z$-pair	&$X$-pair  '\\ 
		$Z$ 	& $Z$-pair 	& $Z$-pair  	&`discard'\\  
		$X$		& $X$-pair 	& `discard'& $X$-pair  \\      
		\hline   
	\end{tabular} 
\end{table}
Finally they will get $Z$-pairs and $X$-pair with different total intensities for the following key mapping and data postprocessing steps.

\subsection{Parameter estimation}
To address data fluctuations, we employ the finite-data-size analysis method, specifically utilizing the Chernoff-Hoeffding method, following the approach outlined in \cite{Zhang2017improved}. For an observed quantity $\chi$, the upper and lower bounds of the underlying expectation value are determined by
\begin{equation}
	\begin{split}
		\mathbb{E}^L(\chi)&=\dfrac{\chi}{1+\delta^L},\\
		\mathbb{E}^U(\chi)&=\dfrac{\chi}{1-\delta^U},
	\end{split}
\end{equation}
where $\delta^{L(U)}$ is the solution to the equations
\begin{equation}\label{eq:deltasolve}
	\begin{split}
		\left[\dfrac{e^{\delta^L}}{(1+\delta^L)^{1+\delta^L}}\right]^{\chi / (1+\delta^L)}&=\dfrac{1}{2}\varepsilon,\\
		\left[\dfrac{e^{-\delta^U}}{(1-\delta^U)^{1-\delta^U}}\right]^{\chi / (1-\delta^U)}&=\dfrac{1}{2}\varepsilon,
	\end{split}
\end{equation}
where $\varepsilon$ represents the security parameter, and its value can be found in Table \ref{table:Fieldparameters}.

With the finite-data-size analysis method in place, Alice and Bob can establish bounds on ${M_{11}^Z}$ and $e^{Z,ph}_{11}$ through the decoy state method by utilizing experimental results under different light intensities.
Firstly, Alice and Bob use $Z$-pairs with different intensity settings to estimate the number of clicked single-photon pairs ${M_{11}^Z}$. Here they use the pairs with intensity $\vec{\mu} = \{(\mu,\mu),(\nu,\nu),(\mu,\nu),(\nu,\mu),(\mu,0),(0,\mu),(\nu,0),(0,\nu),(0,0)\}$ to estimate the lower bound $M^{Z,L}_{11}$ by solving the linear programming problem,
\begin{equation}
	\begin{aligned}
		min & \ M_{11}^Z,    \\
		s.t. &\ \mathbb{E}^L[\mathcal{M}^{\vec{\mu}}] \leq
		\sum_{\vec{k}} \mathrm{Pr}(\vec{\mu}|\vec{k}) M_{\vec{k}}^Z \leq
		\mathbb{E}^U[\mathcal{M}^{\vec{\mu}}],   \\
		& 0 \leq M_{\vec{k}}^Z,    \\
		& \sum_{\vec{k}} M_{\vec{k}}^Z \leq \sum_{\vec{\mu}} \mathcal{M}^{\vec{\mu}},
	\end{aligned}    
\end{equation}
where $\mathbb{E}^L[\mathcal{M}^{\vec{\mu}}]$ and $\mathbb{E}^U[\mathcal{M}^{\vec{\mu}}]$ are results of considering statistical fluctuation by using the finite-data-size analysis method, and $\mathrm{Pr}(\vec{\mu}|\vec{k})$ is the conditional probability that Alice and Bob emit the state with intensity setting $\vec{\mu}$ conditioned on choosing the photon number $\vec{k}$, which is considered to follow the product of two Poisson distribution.

Alice and Bob then use all $X$-pairs with intensity $\vec{\mu} = \{(2\mu,2\mu),(2\nu,2\nu),(2\mu,2\nu),(2\nu,2\mu),
(2\mu,0),(0,2\mu),(2\nu,0),(0,2\nu),(0,0)\}$ for parameter estimation. They estimate the upper bound of single photon phase error rate by solving the following two linear programming problems and get the lower bound of the number of the single photon component, $M_{11}^{X,L}$, and the upper bound of the errors, $E_{11}^{X,U}$, in the $X$-basis,

\begin{equation}\label{eqn-1}
	\begin{aligned}
		min & \ M_{11}^X,   \\
		s.t. &\ \mathbb{E}^L[\mathcal{M}^{\vec{\mu}}] \leq
		\sum_{\vec{k}} \mathrm{Pr}(\vec{\mu}|\vec{k}) M_{\vec{k}}^X \leq
		\mathbb{E}^U[\mathcal{M}^{\vec{\mu}}],   \\
		& 0 \leq M_{\vec{k}}^X,    \\
		& \sum_{\vec{k}} M_{\vec{k}}^X \leq \sum_{\vec{\mu}} \mathcal{M}^{\vec{\mu}},
	\end{aligned}    
\end{equation}
and
\begin{equation}
	\begin{aligned}
		\mathrm{max\quad} &E_{11}^X, \\
		\mathrm{s.t.\quad} & \mathbb{E}^L[\mathcal{M}^{\vec{\mu}}] \leq
		\sum_{\vec{k}} \mathrm{Pr}(\vec{\mu}|\vec{k}) M_{\vec{k}}^X \leq
		\mathbb{E}^U[\mathcal{M}^{\vec{\mu}}],   \\
		& \mathbb{E}^L[\mathcal{E}^{\vec{\mu}}] \leq
		\sum_{\vec{k}} \mathrm{Pr}(\vec{\mu}|\vec{k}) E_{\vec{k}}^X \leq
		\mathbb{E}^U[\mathcal{E}^{\vec{\mu}}],   \\
		& 0 \leq E_{\vec{k}}^X \leq M_{\vec{k}}^X,    \\
		& \sum_{\vec{k}} M_{\vec{k}}^X \leq \sum_{\vec{\mu}} \mathcal{M}^{\vec{\mu}},    \\
		& \sum_{\vec{k}} E_{\vec{k}}^X \leq \sum_{\vec{\mu}} \mathcal{E}^{\vec{\mu}}, \\
		& E_{(i,0)}^X = 0.5M_{(i,0)}^X, \\
		& E_{(0,i)}^X = 0.5M_{(0.i)}^X,
	\end{aligned}
\end{equation}

where $\mathcal{M}^{\vec{\mu}}$ and $M_{\vec{k}}^X$ is the number of $X$-pairs, while $\mathcal{E}^{\vec{\mu}}$ and $E_{\vec{k}}^X$ is the number of error pairs. Moreover, as the pairs with intensity $\vec{\mu} = \{(2\mu,2\mu),(2\nu,2\nu),(2\mu,2\nu),(2\nu,2\mu)$ are sifted by phase, an additional coefficient $(2/D)$ is multiplied to the conditional probability.
With the results of the two bounds, we can further bound the phase error of the single-photon components in the $Z$-basis,
\begin{equation}
	\begin{split}
		e^{Z,ph}_{11}\leq &\dfrac{E_{11}^{X,U}}{M_{11}^{X,L}}\\
		&\equiv e^{Z,ph,U}_{11}.
	\end{split}
\end{equation}
Finally, Alice and Bob substitute $M_{11}^{Z,L}$ and $e^{Z,ph,U}_{11}$ into $M_{11}^{Z}$ and $e^{Z,ph}_{11}$ in Eq.~\eqref{eq:keyR}, obtaining the lower bound on the key number.

\section{Channel stability}\label{app:stable}

\begin{figure}[h]
	\centering
	\includegraphics[width=.7\textwidth]{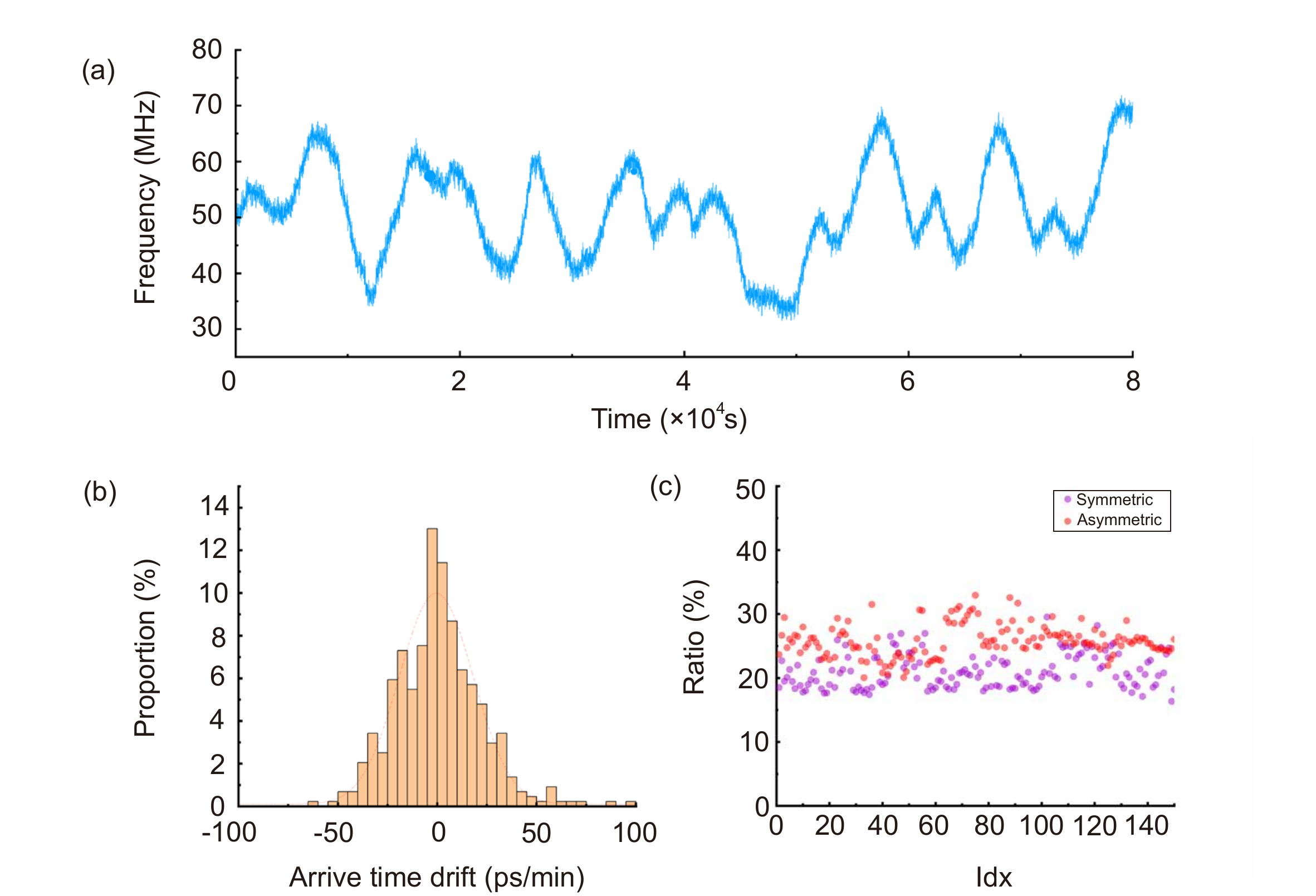}
	\caption{Frequency drift and time drift. (a) The frequency difference drift between two lasers. We measured and recorded the frequency difference by a frequency counter (Keysight 53220A) based on the interference of two lasers which are allocated at two separate sites. The interval between two points is 6s. (b) The photon’s arrival time. We utilize SNSPD3 and SNSPD4 to detect the arriving photon and record the arrival time. (c) The outdoor ratio. We set a detection window in the time domain and discard the data out of the window. Each data point is 30s for the symmetric scenario and 15s for the asymmetric scenario. }	
	\label{fig:freqdrift}
\end{figure}

To assess the stability of the channel, we conducted tests on the long-term frequency difference variations between the two independent lasers at a fiber length of 195.8 km, and the results are shown in Figure \ref{fig:freqdrift}. The results demonstrate that the 2 kHz narrow-linewidth lasers employed in our experiment exhibit considerable stability in real-world environmental conditions.

\begin{figure}[h]
	\centering
	\includegraphics[width=.6\textwidth]{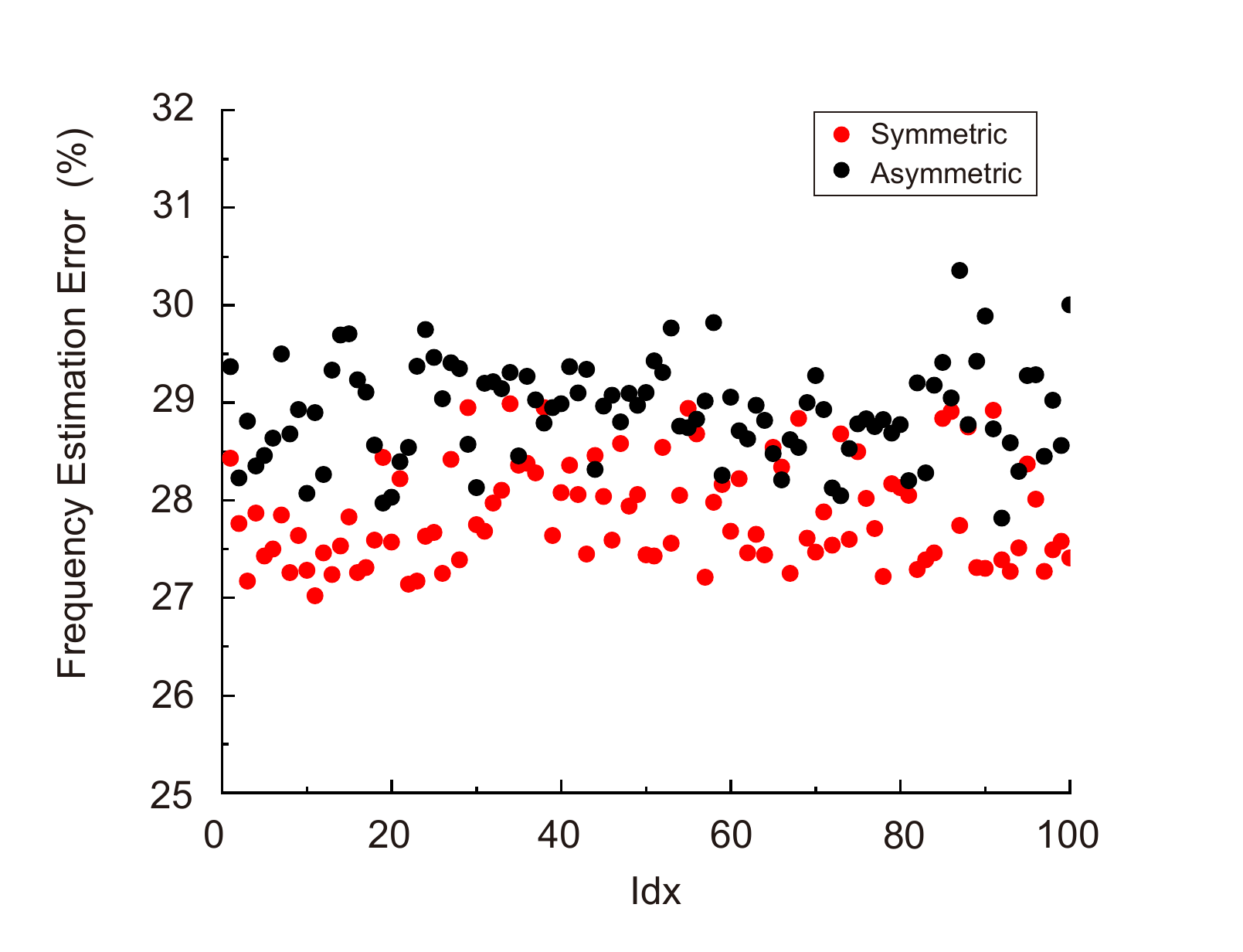}
	\caption{Frequency estimation error under symmetric and asymmetric tests. Each data point represent the effective clicks in 30 seconds. The results indicate that, under long-term operation, the frequency estimation error rates are quite stable for both symmetric and asymmetric scenarios. Although there is a slight increase in error rates in the asymmetric case, it remains acceptable.}	
	\label{fig:freqEstError}
\end{figure}

In Figure \ref{fig:freqEstError}, we also demonstrate the variation of phase estimation error rates over time in both symmetric and asymmetric scenarios. Here, the error rate is defined as the rate of incorrectly predicting the click results in the reference region response when using the phase estimation results. Due to the nature of coherent state interference, there is a 25\% background error rate. These results indicate that the MP scheme can perform phase estimation in asymmetric scenarios without introducing excessively high error rates. In contrast, the twin-field schemes cannot perform phase difference calculation and compensation beyond the coherence length in asymmetric scenarios.

\clearpage

\section{Experimental results and raw data}\label{app:detaildata}

Here, we present the specific parameter settings and results of the experiment in Table \ref{table:Exp-Sim}, comparing them with simulated results. The comparative analysis indicates a consistent agreement between experimental and simulated outcomes, demonstrating that the system operated without additional issues during the experiments. Here, $p_\mu$ denotes the sending probability, $M$ represents the corresponding number of pairs, $EM$ stands for the number of errors in the respective pairs, and $e$ indicates the corresponding error rate. It should be noted that the simulated results show some discrepancies with the experimental results when dealing with pairs with zero intensity. This discrepancy is partly attributed to the typically small number of pairs, leading to greater data fluctuations. Additionally, the numbers of pairs in these cases are often influenced by dark counts. Due to challenges in accurately calculating the dark count rate caused by external light leakage into the optical fiber, the dark count rate we use in simulation tends to be higher.
\begin{table}[h]
	\renewcommand\arraystretch{1.2}
	\caption{Experimental results and the corresponding simulated results.}
	\centering
	\begin{tabular}{p{2cm}|p{2cm}|p{2cm}|p{2cm}|p{2cm}}
		\hline
		\hline
		& \multicolumn{2}{c|}{Symmetry} & \multicolumn{2}{c}{Asymmetry}  \\
		\hline
		& Experiment & Simulation & Experiment & Simulation \\
		\hline
		$ \mu_A $ & 0.3609 & 0.3609 & 0.1480 & 0.1480 \\		
		\hline
		$ \nu_A $ & 0.0360 & 0.0360 & 0.0032 & 0.0032 \\		
		\hline
		$ \mu_B $ & 0.3337 & 0.3337 & 0.6800 & 0.6800 \\		
		\hline
		$ \nu_B $ & 0.0343 & 0.0343 & 0.0541 & 0.0541 \\		
		\hline
		$ p_{\mu_A} $ & 0.25 & 0.25& 0.12 & 0.12\\		
		\hline
		$ p_{\nu_A} $ & 0.25 & 0.25& 0.25 & 0.25\\		
		\hline
		$ p_{\mu_B} $ & 0.25 & 0.25 & 0.25 & 0.25\\		
		\hline
		$ p_{\nu_B} $ & 0.25 & 0.25 & 0.25 & 0.25\\		
		\hline
		
		$	M^Z_{00}	$	&	3	&	0.65	&	184	&	13.14\\
		\hline
		$	M^Z_{0 \mu}	$	&	8955	&	10507.12	&	58189	&	83948.07\\
		\hline
		$	M^Z_{0 \nu }	$	&	948	&	10912.97	&	4830	&	6705.57\\
		\hline
		$	M^Z_{ \mu0}	$	&	9238	&	10903.49	&	63256	&	91145.82\\
		\hline
		$	M^Z_{ \nu 0}	$	&	966	&	1090.77	&	2986	&	4131.26\\
		\hline
		$	M^Z_{ \mu \mu}	$	&	88377993	&	87618992.37	&	294929141	&	291130847.10\\
		\hline
		$	M^Z_{ \mu \nu }	$	&	9078513	&	9012761.89	&	23581217	&	23296775.26\\
		\hline
		$	M^Z_{ \nu  \mu}	$	&	8833952	&	8762432.97	&	13334537	&	13228325.89\\
		\hline
		$	M^Z_{ \nu  \nu }	$	&	908463	&	901330.56	&	1064193	&	1058543.28\\
		\hline
		$	E^Z_{ \mu \mu}	$	&	9020	&	10698.80	&	42930	&	61383.52\\
		\hline
		$	E^Z_{ \mu \nu }	$	&	5021	&	5995.14	&	32318	&	46833.55\\
		\hline
		$	E^Z_{ \nu  \mu}	$	&	4998	&	5797.53	&	24738	&	35365.14\\
		\hline
		$	E^Z_{ \nu  \nu }	$	&	947	&	1085.05	&	3376	&	4720.72\\
		\hline
		$	e^Z_{ \mu \mu}	$	&	$1.021 \times 10^{-4}$	&	$1.221 \times 10^{-4}$	&	$1.456  \times 10^{-4}$ 	&	$2.108 \times 10^{-4}$\\
		\hline
		$	e^Z_{ \mu \nu }	$	&	$5.531 \times 10^{-4}$	&	$6.652 \times 10^{-4}$	&	$1.370 \times 10^{-3}$	&	$2.010 \times 10^{-3}$\\
		\hline
		$	e^Z_{ \nu  \mu}	$	&	$5.658 \times 10^{-4}$	&	$6.616 \times 10^{-4}$	&	$1.855 \times 10^{-3}$	&	$2.673 \times 10^{-3}$\\
		\hline
		$	e^Z_{ \nu  \nu }	$	&	$1.042 \times 10^{-3}$	&	$1.204 \times 10^{-3}$	&	$3.172 \times 10^{-3}$	&	$4.460 \times 10^{-3}$\\
		\hline
		$	M^X_{00}	$	&	3	&	0.65	&	184	&	13.14\\
		\hline
		$	M^X_{02 \mu }	$	&	43376782	&	42175066.32	&	138310497	&	134041912.30\\
		\hline
		$	M^X_{02 \nu }	$	&	459106	&	445762.12	&	881145	&	855246.02\\
		\hline
		$	M^X_{2 \mu 0}	$	&	46652989	&	45496153.13	&	162706031	&	158012951.20\\
		\hline
		$	M^X_{2 \nu 0}	$	&	465156	&	454526.0446	&	331330	&	324626.4742\\
		\hline
		$	M^X_{2 \mu 2 \mu }	$	&	5549824	&	5465984.76	&	9113959	&	8958448.24\\
		\hline
		$	M^X_{2 \mu 2 \nu }	$	&	1757516	&	1716316.32	&	5389093	&	5214867.43\\
		\hline
		$	M^X_{2 \nu 2 \mu }	$	&	1642605	&	1605034.24	&	3058876	&	2973581.79\\
		\hline
		$	M^X_{2 \nu 2 \nu }	$	&	57045	&	56221.14	&	54259	&	52984.05\\
		\hline
		$	EM^X_{00}	$	&	1.5	&	0.33	&	92	&	6.57\\
		\hline
		$	EM^X_{02 \mu }	$	&	21607575	&	21087533.16	&	69704510	&	67020956.14\\
		\hline
		$	EM^X_{02 \nu }	$	&	229101	&	222881.06	&	444240	&	427623.01\\
		\hline
		$	EM^X_{2 \mu 0}	$	&	23313611	&	22748076.56	&	81983904	&	79006475.62\\
		\hline
		$	EM^X_{2 \nu 0}	$	&	231874	&	227263.02	&	166966	&	162313.24\\
		\hline
		$	EM^X_{2 \mu 2 \mu }	$	&	1882646	&	1854205.53	&	3468649	&	3409463.72\\
		\hline
		$	EM^X_{2 \mu 2 \nu }	$	&	788350	&	769869.51	&	2608963	&	2524617.07\\
		\hline
		$	EM^X_{2 \nu 2 \mu }	$	&	732276	&	715526.89	&	1427742	&	1387930.60\\
		\hline
		$	EM^X_{2 \nu 2 \nu }	$	&	19683	&	19398.73	&	19282	&	18828.92\\
		\hline
		$	e^X_{2 \mu 2 \mu }	$	&	0.3392	&	0.3392	&	0.3806	&	0.3806\\
		\hline
		$	e^X_{2 \mu 2 \nu }	$	&	0.4486	&	0.4486	&	0.4841	&	0.4841\\
		\hline
		$	e^X_{2 \nu 2 \mu }	$	&	0.4458	&	0.4458	&	0.4668	&	0.4668\\
		\hline
		$	e^X_{2 \nu 2 \nu }	$	&	0.3450	&	0.3450	&	0.3554	&	0.3554\\
		\hline
		
		\hline
	\end{tabular}	
	\label{table:Exp-Sim}
\end{table}

Furthermore, Table \ref{table:ExpProcess} presents the processing results for the experimental data, including the lower bounds on the estimated single-photon component in different $Z$-pairs $M^{\vec{\mu},L}_{11}$, the upper bounds on the phase error rate $e^{Z,ph}_{11}$, as well as the total number of key bits $MR$ and key rate $R$.
\begin{table}[h]
	\renewcommand\arraystretch{1.2}
	\caption{Results of experimental data processing}
	\centering
	\begin{tabular}{l|p{2cm}|p{2cm}}
		\hline
		\hline
		& Symmetry & Asymmetry \\
		\hline
		$	M^{(\mu,\mu),L}_{11}	$	&	43019763	&	124841146\\
		\hline
		$	M^{(\mu,\nu),L}_{11}	$	&	5959809	&	18596822\\
		\hline
		$	M^{(\nu,\mu),L}_{11}	$	&	5944932	&	6510179\\
		\hline
		$	M^{(\nu,\nu),L}_{11}	$	&	823590	&	969782\\
		\hline
		$e^{Z,ph}_{11}	$	&	0.3131	&	0.3152		\\
		\hline
		$	MR	$	&	5477255		&	13899131		\\
		\hline
		$	R \text{(bit per pair)}	$	&	$5.47\times10^{-6}$		&	$1.39\times10^{-5}$	\\
		\hline
		$	R \text{(bit per second)}	$	&	1217.17	&	3088.69	\\
		\hline
		\hline
	\end{tabular}	
	\label{table:ExpProcess}
\end{table}

\end{document}